\documentclass[letterpaper]{article} 
\usepackage{aaai25}  
\usepackage{times}  
\usepackage{helvet}  
\usepackage{courier}  
\usepackage[hyphens]{url}  
\usepackage{graphicx} 
\usepackage{natbib}  
\usepackage{caption} 
\usepackage{algorithm}
\usepackage{algpseudocode}
\usepackage{newfloat}
\usepackage{listings}
\usepackage{amssymb}
\usepackage{amsmath}
\usepackage{subfig}
\usepackage{booktabs}
\DeclareCaptionStyle{ruled}{labelfont=normalfont,labelsep=colon,strut=off} 
\floatstyle{ruled}
\newfloat{listing}{tb}{lst}{}
\floatname{listing}{Listing}
\title{HLoRA: Efficient Federated Learning System for LLM \\ Heterogeneous Fine-Tuning}
\author{
    Qianli Liu\textsuperscript{\rm 1},
    Zhaorui Zhang\textsuperscript{\rm 2},
    Xin Yao\textsuperscript{\rm 3},
    Benben Liu\textsuperscript{\rm 3}
}
\affiliations{
    \textsuperscript{\rm 1}Hong Kong University of Science and Technology\\
    \textsuperscript{\rm 2}Hong Kong Polytechnic University\\

    \textsuperscript{\rm 3}The University of Hong Kong\\
}

\usepackage{bibentry}

\begin{document}

\maketitle

\begin{abstract}
Federated learning systems have been identified as an efficient approach to scaling distributed model training with a large amount of participants or data owners while guaranteeing data privacy. To apply the current most popular pre-trained large language models to other domains with data privacy guarantee requirements, existing works propose fine-tuning the pre-trained large language models in federated learning environments across data owners using the parameter efficient fine-tuning approaches, LoRA. To address the resource and data heterogeneous issues for the participants, previous works adopted heterogeneous LoRA using different ranks for different clients and pending their rank, which brings bias for the parameter aggregation. 

To address this issue, we propose HLoRA, an efficient federated learning system utilizing a modified LoRA approach that incorporates rank heterogeneity to optimize communication and computational efficiency. Experimental results, conducted using the Microsoft Research Paraphrase Corpus (MRPC), Quora Question Pairs (QQP) and Recognizing Textual Entailment (RTE), within the Plato federated learning framework, demonstrate that our method not only reduces resource demands but also outperforms traditional LoRA applications in terms of convergence speed and final model accuracy. This study shows that our approach can significantly improve the practical deployment of federated LLM fine-tuning, particularly in environments with diverse client resources.
\end{abstract}

%

\section{Introduction}
In recent years, large language models (LLMs) have achieved a great breakthrough \cite{touvron2023llama, openai2023gpt, zhang2022opt, zeng2022glm} and have been widely used in many domains, including advanced ChatBots for diverse writing tasks \cite{chatgpt}, and as a component of multi-modal systems \cite{driess2023palm,anil2023palm,chowdhery2023palm}, text/image generation with prompts, language translation, solving math problems. Lots of pre-trained large language models that are trained based on the public dataset, such as data collected through the Internet, have been open-sourced and achieved great success for general tasks. Recent progress in large language modeling has relied heavily on unsupervised training on vast amounts of human-generated text, primarily sourced from the web or curated corpora \cite{zhao2023survey}. The emerging largest datasets of human-generated public text data, including Refined Web, C4, and RedPajama, contain tens of trillions of words collected from billions of web pages \cite{togetherai}. To achieve a higher accuracy for the large language models, the demand for public human text data is likely to continue growing. To scale up the large language models and train them efficiently, they are typically trained according to the neural scaling laws \cite{hoffmann2022training}. Such relationships indicate that increasing the size of the training datasets is essential for efficiently improving the performance of the LLMs. However, according to the estimation of the data stocks \cite{villalobosposition}, the high-quality public data will be used up within a few years in the future \cite{longpre2024consent}. 

As a consequence, fine-tuning the large language models in specified domains based on private data owned by different organizations or institutes, such as government and hospitals, has become a new direction to enhance the development of the large language models. This is also benefited by the surprising zero/few-shot learning capabilities of the emerging foundation models (LLMs). Existing LLMs, such as GPT \cite{achiam2023gpt, brown2020language} and PaLM series \cite{driess2023palm}, are trained on a massive variety of data (mostly unlabeled) with parameters ranging up to hundreds of billions in size, making it capable of being applied to different domains with just a few additional training rounds (such as fine-tuning) on the targeted dataset \cite{cho2023heterogeneous}. 

Federated learning systems have been identified as an efficient approach to scaling distributed model training with a large number of participants or data owners while guaranteeing the privacy of the training data \cite{xu2024fedfa, zhang2022mipd, zhang2021sapus, zhang2022momentum, zhang2025cllora}. Therefore, fine-tuning the pre-trained large language models in federated learning environments becomes the best choice for applying the pre-trained emerging LLMs in specified domains based on private data. This adaptation process utilizes task-specific data to tailor a model, enabling it to perform optimally across various applications\cite{howard2018universal}. The colossal size of the emerging high-accuracy LLMs, however, requires a large amount of resources for directly fine-tuning their entire parameter space. To tackle this issue, some recent works have been proposed for parameter-efficient fine-tuning (PEFT) of the LLMs, such as prompt tuning \cite{lester2021power}, utilizing adapters \cite{houlsby2019parameter}, or low-rank adaptation (LoRA) of the original models \cite{hu2021lora}, which freezes the original pre-trained parameters of the LLMs and train only additional, smaller part of parameters instead. Such an approach can not only reduce the computation overhead during the training procedure but also reduce the communication overhead in distributed training environments since it only transmits part of the trainable parameters between clients and servers.

In this work, we investigate a simple, scalable technique for applying parameter-efficient fine-tuning (LoRA) \cite{hu2021lora} to the existing pre-trained large language models in heterogeneous federated learning environments. However, this is non-trivial due to the distributed and heterogeneous features of federated learning systems. We identify two challenges that apply the LoRA in heterogeneous federated learning systems.

\textit{Firstly,} due to the heterogeneity of the resources and data for different clients in heterogeneous federated learning systems, applying the same rank of LoRA approach for all clients is inefficient. Adaptively adjusting the rank of the LoRA for different clients is an efficient method to address this issue. However, this brings challenges for parameter aggregation on the server for the parameter that is collected from clients with different ranks. Simply pending for parameters with different ranks involving bias for the parameter aggregation \cite{cho2023heterogeneous}. How to efficiently aggregate the parameters that are collected from different clients with different ranks is challenging.

\textit{Secondly,} after aggregation on the server for the parameters that are collected from clients, the server needs to decompose the parameters based on the LoRA approach and assign a suitable rank for different clients. Estimating a suitable rank for different clients is also challenging.

To address the above challenges, we propose \textit{HLoRA}, an efficient federated learning system for the fine-tuning of large language models in heterogeneous environments. It can achieve better performance in heterogeneous data environments without increasing communication and computation costs. \textit{HLoRA} allows different clients to adopt different ranks during the fine-tuning procedure by reconstructing the original parameter matrix and then decomposing them. For each communication round, the server collects the two LoRA matrices and multiplies them to reconstruct the original parameter matrix. Then, the average value of the reconstructed parameter matrix will be calculated on the server. Finally, the server will decompose the updated parameters into two LoRA matrices according to the different computation resources and data for different clients and broadcast them to the clients. 

We summarize our contributions as follows:

\begin{itemize}
     \item We give an in-depth analysis and explore the inconsistencies that arise from the simple and direct apply the parameter-efficient fine-tuning approach "LoRA" in heterogeneous federated learning environments. We also provide an explanation of the potential reasons for performance degradation that is caused by integrating the LoRA into heterogeneous federated learning systems.
     \item We propose an efficient approach to fine-tune the large language models based on the parameter-efficient fine-tuning method "LoRA" in heterogeneous federated learning environments that allows different clients to adopt different ranks of the LoRA approach, called \textit{HLoRA}. \textit{HLoRA} reconstruct the original parameter matrix by two LoRA matrices before aggregation to avoid the unnecessary pending for heterogeneous LoRA matrices. After parameter updating, \textit{HLoRA} decomposes the updated parameter matrix into two LoRA matrices according to the different resources and data for different clients and broadcasts them to the clients.  
    \item We perform a comprehensive evaluation of \textit{HLoRA} on the most popular large language models Roberta-Large based on various datasets, including  Microsoft Research Paraphrase Corpus (MRPC), Quora Question Pairs (QQP) and Recognizing Textual Entailment (RTE) in Non-IID scenarios, which are often used in heterogeneous federated learning environments. Evaluation results show that \textit{HLoRA} outperforms baseline for both the model accuracy and training rounds that achieve the target accuracy by up to $1.1\times$.  
    
\end{itemize}

\begin{figure}[t]
    \centering
    \includegraphics[width=0.95\linewidth]{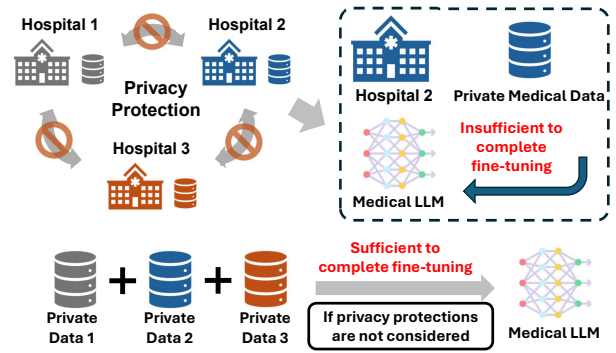}
    \caption{For illustration, consider a consortium of three hospitals aiming to develop an LLM for medical diagnostics. Each entity's data, while valuable, is insufficient in isolation. United, their data could revolutionize medical LLM training. Alas, stringent data privacy laws often thwart such synergistic endeavors, exacerbating the dual challenges of data paucity and privacy preservation.}
    \label{fig:fl}
    \vspace{-8pt}
\end{figure}

\section{Background and Related Works}
\subsection{Parameter-Efficient Fine Tuning (PEFT) for Large Language Models}

Large language models have achieved a great breakthrough in many domains in recent years due to the rapid increase of their network size and training data set. However, training large language models based on a large volume of training data sets becomes very expensive. To address this issue, emerging parameter-efficient fine-tuning strategies are proposed. These techniques typically introduce a minimal number of additional trainable parameters to enhance model performance while maintaining the majority of pre-trained parameters in a frozen state. 
Parameter-efficient fine-tuning strategies include the integration of trainable neural modules, known as adapters \cite{houlsby2019parameter,hanParameterEfficientFineTuningLarge2024,houlsbyParameterEfficientTransferLearning2019}, into each layer of the network. These modules encapsulate the task-specific enhancements in significantly smaller dimensions than the original model parameters. Other approaches, such as prefix-tuning \cite{li2021prefix} and prompt-tuning \cite{lester2021power}, extend the model by appending trainable dimensions to the inputs or hidden layers, thereby modifying the initial conditions or processing pathways of the network. Another innovative PEFT method involves the use of low-rank matrices to approximate \cite{hu2021lora} or re-parameterize pre-trained weight matrices, which is a technique often referred to as rank-grouped parameterization (RGP) \cite{yu2021large}. 

Among various parameter-efficient fine-tuning strategies, LoRA \cite{hu2021lora} is particularly notable as it requires tuning fewer than 1\% of the parameters involved in a comprehensive fine-tuning process yet delivers performance that is competitive across various downstream tasks. Recent studies \cite{he2021towards,chavan2023one} have also explored the development of generalized methods that aim to unify these diverse PEFT approaches. These unified frameworks are designed to streamline the application of PEFT methods, facilitating their adoption in practical settings where model efficiency and adaptability are critical.


\subsection{Large Language Models Fine-Tuning in Federated Learning Environments} 

Fine-tuned large language models (LLMs) have increasingly become integral to applications across various domains, though the fine-tuning process often relies on large-scale, domain-specific datasets \cite{luFederatedLearningNonIID2024,mammenFederatedLearningOpportunities2021,maoUniPELTUnifiedFramework2022}. Typically, these datasets are distributed among multiple stakeholders or data owners across different countries or regions subject to national policies restricting data transfer across regions or countries. Each data owner possesses only a fraction of the data required for effective model training, and direct data sharing is frequently restricted due to privacy concerns \cite{hsuMeasuringEffectsNonIdentical2019, liFederatedOptimizationHeterogeneous, liPrefixTuningOptimizingContinuous2021}. Federated learning (FL) \cite{fl01} offers a promising solution by enabling collaborative model tuning without the need to exchange raw data directly. This method involves stakeholders sharing their local model updates, thereby collectively enhancing the performance of the large language models. 

For instance, the introduction of FedBERT \cite{tian2022fedbert} illustrates the application of federated pre-training on the BERT model. Unlike traditional machine learning models, the substantial size of large language models necessitates significant computational and communication resources for cross-party interactions during training procedures in federated learning environments. Recent study \cite{zhang2022federated} has explored the integration of parameter-efficient fine-tuning with federated learning systems, with multiple studies examining parameter-efficient fine-tuning within this context. Recently, the newly proposed FederatedScope-LLM framework \cite{kuang2023federatedscope} supports fine-tuning the large language models in federated learning environments, which is proposed to address the challenges caused by data heterogeneity, which is a major challenge to apply the parameter-efficient fine-tuning algorithms in federated learning systems. There are also lots of lossy compression approaches that aim to compress the gradient and parameters to reduce the communication overhead \cite{di2024survey,huang2023c,huang2024optimized,huang2025zccl}.

There are also a few works that focus on the utilization of LoRA in federated learning environments. For example, some research has assessed the importance of initialization for LoRA modules \cite{sheng2023s}, proposing that these modules be trained via federated learning followed by singular value decomposition (SVD) to achieve effective initial configurations. However, these approaches do not modify the training process of LoRA to accommodate the diverse system capabilities of different devices. Another study \cite{yi2023fedlora} has investigated LoRA in the context of personalized federated learning, but this also did not adopt the LoRA methodology itself beyond its application to personalization and did not address the heterogeneous problem. The study \cite{cho2023heterogeneous} proposes heterogeneous LoRA that can apply different rank LoRA modules to different clients via utilizing zero-padding and truncation for the aggregation and distribution of the heterogeneous size LoRA modules. However, this work \cite{cho2023heterogeneous} brings bias for the parameter aggregation (introduced in the following sections).

Our research introduces an innovative federated learning system for large language models fine-tuning to cater to system and data heterogeneity. It can achieve better performance in heterogeneous data environments without increasing communication and computation costs.

\begin{figure*}[th]
    \centering
    \includegraphics[width=\textwidth]{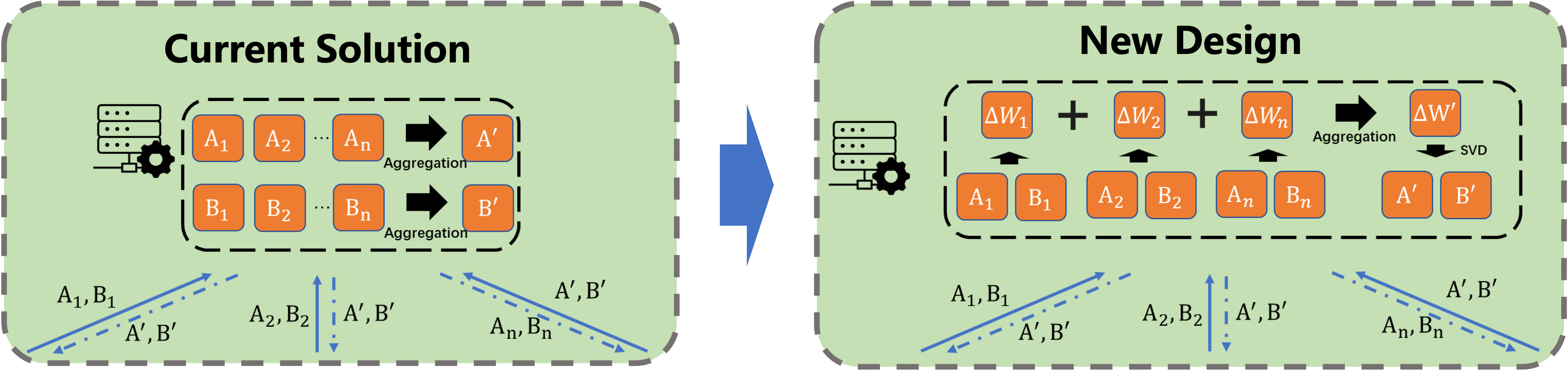}
    \caption{Compared to the direct application of LoRa, our design reconstructs the weight matrix to achieve the optimal effect of aggregating the weights, and at the same time can aggregate the heterogeneous rank between clients}
    \label{fig:fl}
    \vspace{-10pt}
\end{figure*}

\section{Methodology and Design of \textit{HLoRA}}

In this section, we first introduce the naive implementation (homogeneous LoRA), which simply and directly integrates the LoRA into the federated learning systems and lets the ranks of the LoRA for all clients be the same. Then, we identify some limitations of homogeneous LoRA in heterogeneous federated learning systems. Finally, we propose our proposed \textit{HLoRA}, which allows different clients to use different LoRA ranks during the fine-tuning procedure.

\subsection{Naïve Implementation: Homogeneous LoRA}
In the federated learning scenario, the application of the LoRA approach necessitates a distributed computational framework where multiple clients collaboratively train a shared model while maintaining data locality. This section details the process of deploying LoRA under federated learning environments, complemented by a pseudo-code representation of the key steps.

For a pre-trained LLM weight matrix $W_0\in \mathbb{R}^{d\times k}$, standard LoRA method under the centralized training environments uses two low-rank adaptors to constrain its update $W_0+\Delta W=W_0+BA$, where $B\in \mathbb{R}^{d\times r}, A\in \mathbb{R}^{r\times k}$, and the rank $r \ll \min(d,k)$. Only $A$ and $B$ are trainable during the training procedure, while $W_0$ is fixed and receives no gradient updates.

Under the federated learning environments, suppose there are \(K\) clients, each with a local copy of the adaptors \(B_k\) and \(A_k\) for \(k \in \{1, 2, \dots, K\}\). Each client trains and updates its local adaptors \(B_k\) and \(A_k\) using their own data. Subsequently, a central server aggregates these updates to form a global \(B' = \sum_{k=1}^K \eta_k B_k\) and \(A' = \sum_{k=1}^K \eta_k A_k\) aggregation, where \(\eta_k\) represents the weight of the \(k\)-th client's contribution to the overall update. This method allows the model to benefit from diverse data sources while keeping the computational efficiency of low-rank updates. We list the key steps in Algorithm \ref{alg:federated_lora}.

\begin{algorithm}
\caption{Federated Learning with LoRA (Naive) }
\label{alg:federated_lora}
\begin{algorithmic}[1]
\State \textbf{Input:} Initial weight matrix \(W_0\), number of clients \(K\)
\State \textbf{Output:} Updated global adaptors \(B',A'\)

\State Initialize \(B_k\) and \(A_k\) for each client \(k\)
\State Distribute initial model \(W_0\) to each client
\For{each training round}
    \For{each client \(k\) in parallel}
        \State Update \(B_k = B'\) and \(A_k=A'\) 
        \State Train \(B_k\) and \(A_k\) locally on client \(k\)'s data
        \State Upload \(B_k,A_k\) to the server
    \EndFor
    \State Aggregate updates at server: 
    \State \(B' = \sum_{k=1}^K \frac{n_k}{n} B_k\) 
    \State \(A' = \sum_{k=1}^K \frac{n_k}{n} A_k\)
    \State Distribute updated \(B',A'\) to all clients
\EndFor
\State \textbf{return} \(B',A'\)

\end{algorithmic}
\end{algorithm}

\subsection{Limitations of the Naive Implementation for the LoRA in Federated Learning Environments}
\label{parameters_bias}

The naive implementation of the Low-rank Adaptation (LoRA) method in the federated learning environments, as described in the above sections, simplifies several key aspects of practical deployment. While this approach benefits from computational efficiency and reduced communication costs, it also introduces potential issues that could affect the model's performance and fairness across clients. We identify two limitations as the following:

\subsubsection{Bias Introduction Through Parameter Aggregation Procedure.} The naive implementation approach aggregates the LoRA matrices of \(B_k\) and \(A_k\) from each client before updating the global model. This approach deviates from the traditional federated averaging (FedAvg) algorithm, where model parameters are averaged before aggregation. The difference in these approaches can introduce biases in the model updates, as shown in the Fig. \ref{eq:bias_ab}:

\begin{equation}
    W' = \sum_{k=1}^K \frac{n_k}{n} B_k \sum_{k=1}^K \frac{n_k}{n} A_k \ne \sum_{k=1}^K \frac{n_k}{n} B_k A_k 
    \label{eq:bias_ab}
\end{equation}

\subsubsection{Client Heterogeneity and Rank Diversity.} The naive implementation assumes that all clients fine-tuning the pre-trained models using adaptors \(B_k\) and \(A_k\) based on the same rank \(r\). However, in practical federated learning environments, client heterogeneity—variations in data volume, computational power, and privacy requirements—often means that different clients might benefit from or be capable of supporting different ranks. The current aggregation method does not accommodate varying ranks, as it strictly requires uniformity in the dimensions of \(B_k\) and \(A_k\) across all clients. This lack of flexibility can lead to suboptimal learning or participation barriers for less capable clients.

These issues suggest that while the naive implementation of the LoRA in federated learning environments offers a starting point for incorporating low-rank adaptations into federated learning systems, further refinements are necessary to address biases and client heterogeneity effectively. The following subsection will explore potential enhancements to this basic framework to overcome these limitations.

\subsection{The Design of \textit{HLoRA}}
We propose a novel aggregation method for different low-rank adaptors on the server. The method is not restricted to any rank range and maintains high performance despite client rank heterogeneity. For the sake of formality, in this paper, we specify that each client has a rank denoted by $r_k$. 
Our proposed joint fine-tuned heterogeneous rank LoRA module has two main steps: 
(1) model parameters reconstruction and aggregate on the server
(2) the aggregated model parameters are decomposed and assigned low-rank matrices of specified sizes according to the client's requirements.
We will describe each step in detail in the subsequent paragraphs of this section.


\subsubsection{Model Parameters Reconstruction and Aggregation on the Server.} 

In the initial step, we address the aggregation process by directly combining the products of the \(B_k\) and \(A_k\) matrices that are collected from each client on the server. This method contrasts sharply with the naive approach, where the products of matrices were aggregated separately, which led to a skewed representation of client contributions. Our proposed approach integrates the client-specific adaptations directly as the following formula (\ref{our_ab}):

\begin{equation}
 W' = \sum_{k=1}^K \frac{n_k}{n} (B_k A_k)
\label{our_ab}
\end{equation}

This equation ensures that each client’s adaptors contribute as unified entities, preserving the unique data characteristics of each dataset. This aggregation not only eliminates the bias introduced by separate aggregations but also simplifies the update process to the global model by treating each client’s contribution holistically, thereby enhancing the representativeness and robustness of the model.

\subsubsection{Updated Model Parameters Decomposition and Assignment of Ranks} 

Upon aggregating the global model parameters matrix \( W'\), we apply a matrix factorization technique such as Singular Value Decomposition (SVD) to decompose it into its constituent elements. This decomposition is crucial as it allows us to distill and retain the most informative features of the aggregated matrix, which are paramount for reconstructing the low-rank matrices that are specifically tailored to the capabilities and needs of each client, which is shown in the following formula (\ref{decompose}),

\begin{equation}
     W' = U \Sigma V^T \rightarrow B'_k = U_{r_k}, \quad A'_k = \Sigma_{r_k} V^T_{r_k}
     \label{decompose}
\end{equation}

The matrices \(U_{r_k}\), \(\Sigma_{r_k}\), and \(V^T_{r_k}\) represent the truncated versions of \(U\), \(\Sigma\), and \(V^T\), respectively, including only the top \(r_k\) singular values and vectors. This selective truncation ensures that the adaptors \(B'_k\) and \(A'_k\) are optimized for performance but scaled according to each client's computational and data handling capacity. The rank \(r_k\) is predetermined based on a balance between computational feasibility and the necessity to capture sufficient data characteristics, ensuring that each client receives a model that is both manageable and effective. This tailored approach not only improves the efficiency of data representation but also enhances the overall adaptability of the federated learning system to diverse client environments.

We list the key steps of our \textit{HLoRA} as the following:

\begin{enumerate}
    \item \textbf{Local Training:} Each client \(k\), where \(k \in \{1, 2, \dots, K\}\), independently train the model locally and calculates the updates for \(B_k\) and \(A_k\) using their local datasets. This step involves optimizing the local models to best fit the data available at each node, subject to the constraint that the updates must remain within the low-rank structure specified by \(B\) and \(A\).
    
    \item \textbf{Uploading LoRA Matrics:} After local training, each client then uploads $B_k$ and $A_k$ to a central server. This method reduces the dimensionality of the data that needs to be communicated, aligning with the privacy and efficiency goals of federated learning. For our proposed \textit{HLoRA}, the ranks of $B_k$ and $A_k$ for different clients and even different transformer layers can be different.

    
    \item \textbf{Aggregation at the Server:} Upon receiving the updates from all clients, the central server performs an aggregation step. The server calculates the product $W_k = B_k \cdot A_k$ for each client. Then the server calculates the average value of the aggregated parameters $W' = \sum_{k=1}^K \frac{n_k}{n} B_k \cdot A_k$, the same with FedAvg, which synthesizes the contributions from all participating clients into an update matrix for the global model.
    
 
    \item \textbf{Model Updating:} The aggregated update $W' = \sum_{k=1}^K \frac{n_k}{n} B_k \cdot A_k$ is then used to update the global model's parameters matrix, resulting in the new global model $W'$. This updated adaptor is subsequently redistributed to all clients, ensuring that each client starts the next round of training with the updated global adaptor. After the model updating, the server will decompose the updated model parameters $W'$ as two LoRA matrics according to the different computation and data resources of each client, and send them to each client.
\end{enumerate}

\begin{figure*}[t!] 
    \centering
    \subfloat[Naive implementation and Homogeneous \textit{HLoRA} (MRPC)]{\includegraphics[width=0.5\textwidth, height=0.32\linewidth]{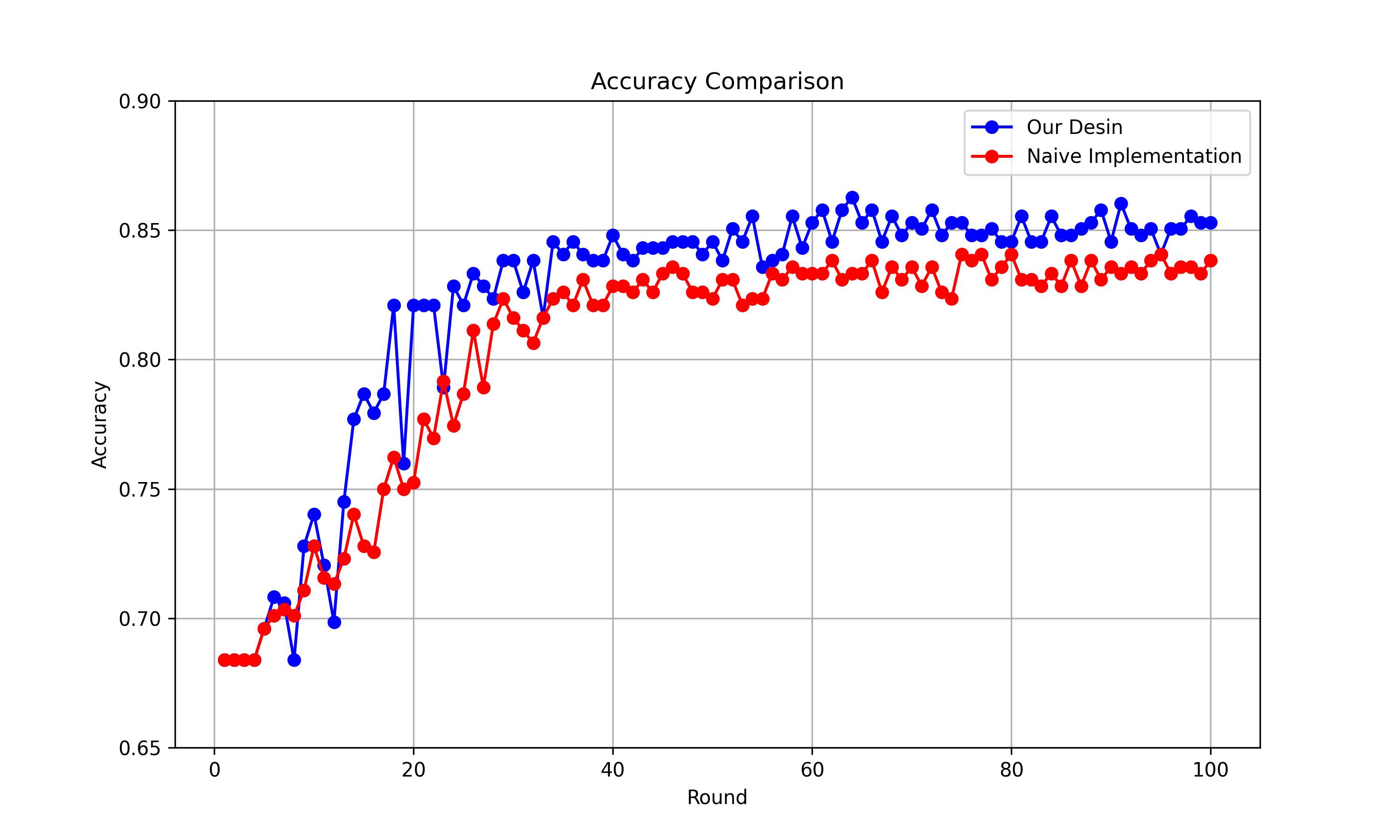}
    \label{fig:mrpc_1}}    
        \subfloat[Homogeneous \textit{HLoRA} and Heterogeneous \textit{HLoRA}(MRPC)]{\includegraphics[width=0.5\textwidth, height=0.32\linewidth]{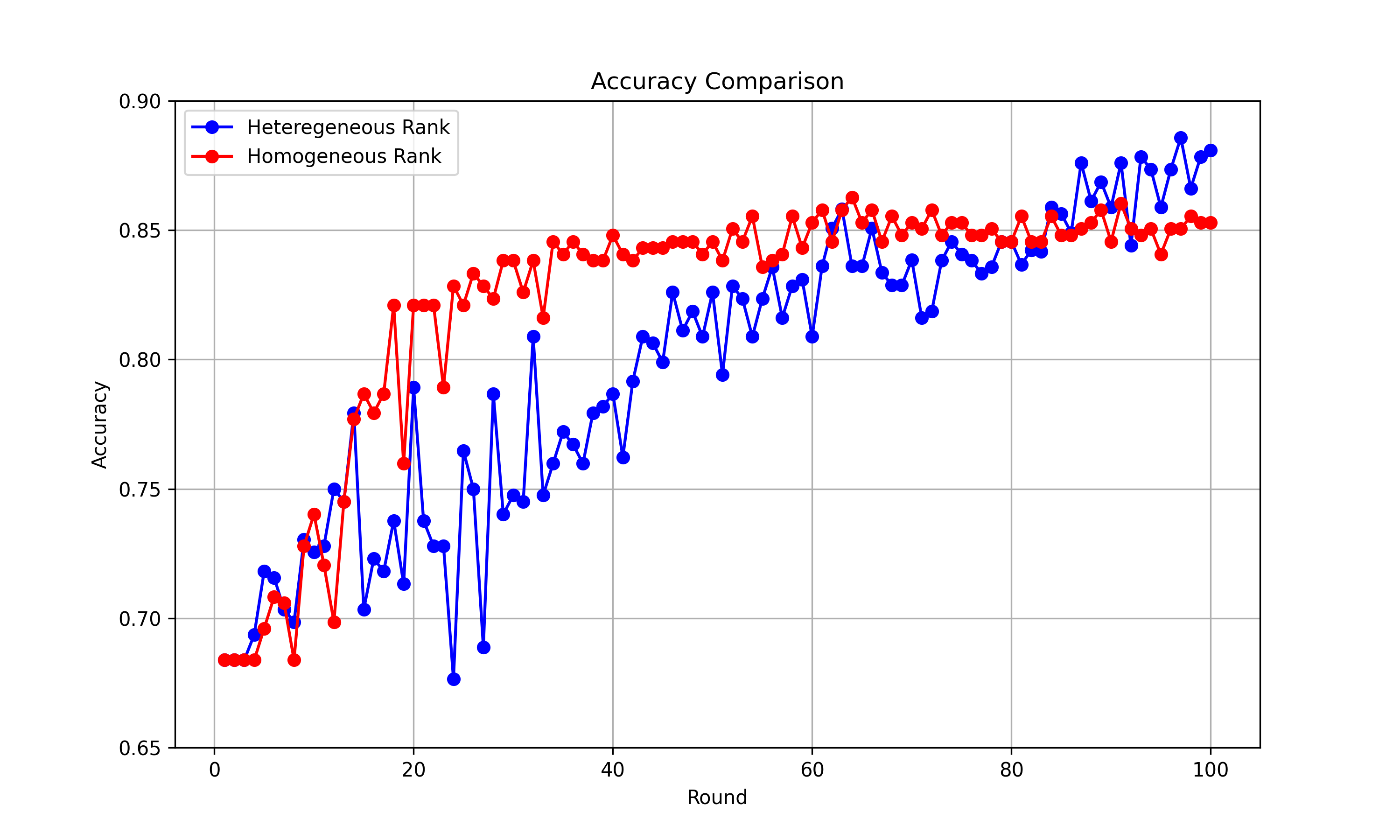}\label{fig:mrpc_2}}  \\ 
        
        \vspace{-13pt}
        
    \subfloat[Naive implementation and Homogeneous \textit{HLoRA} (RTE)]{\includegraphics[width=0.5\textwidth, height=0.32\linewidth]{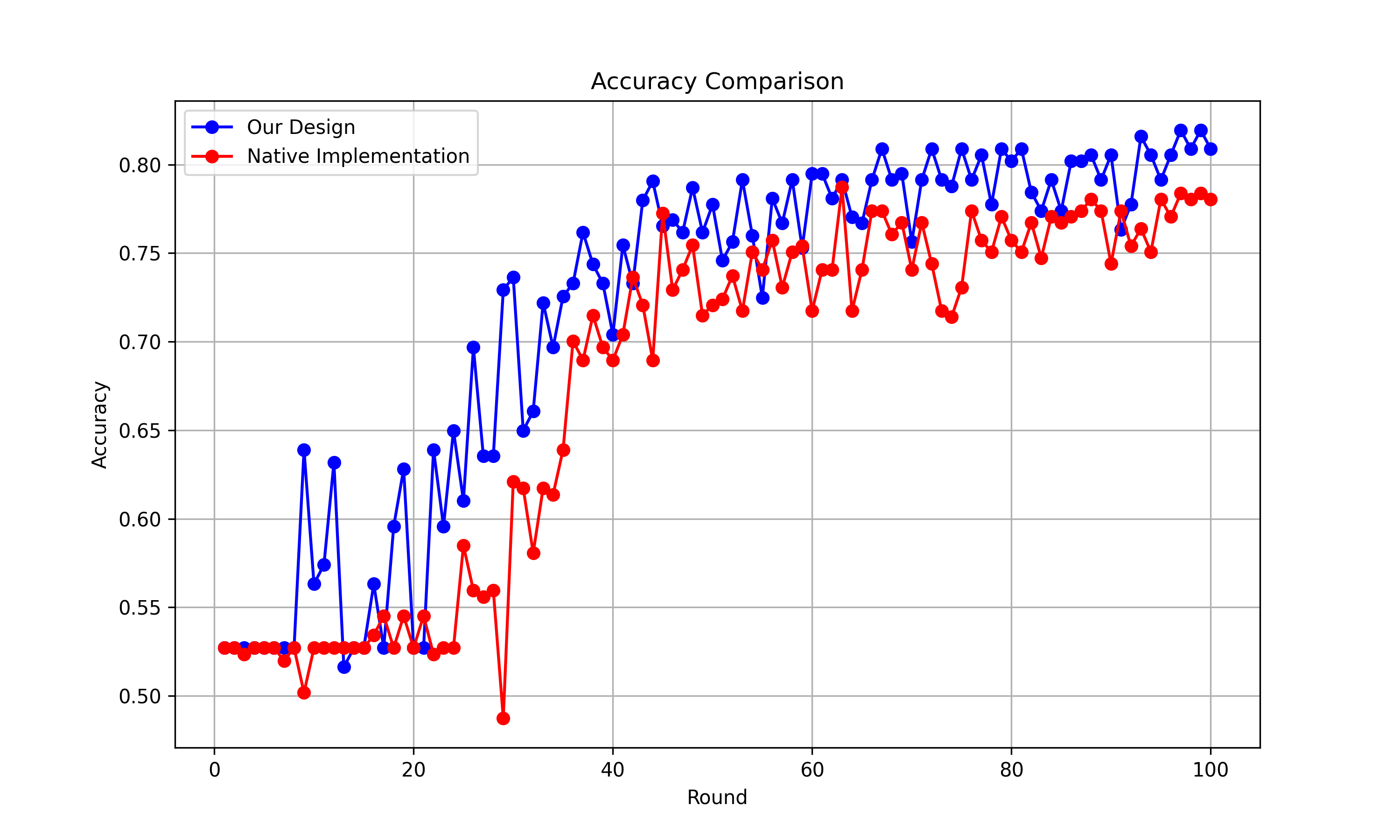}\label{fig:rte_1}} 
        \subfloat[Homogeneous \textit{HLoRA} and Heterogeneous \textit{HLoRA}(RTE)]{\includegraphics[width=0.5\textwidth, height=0.32\linewidth]{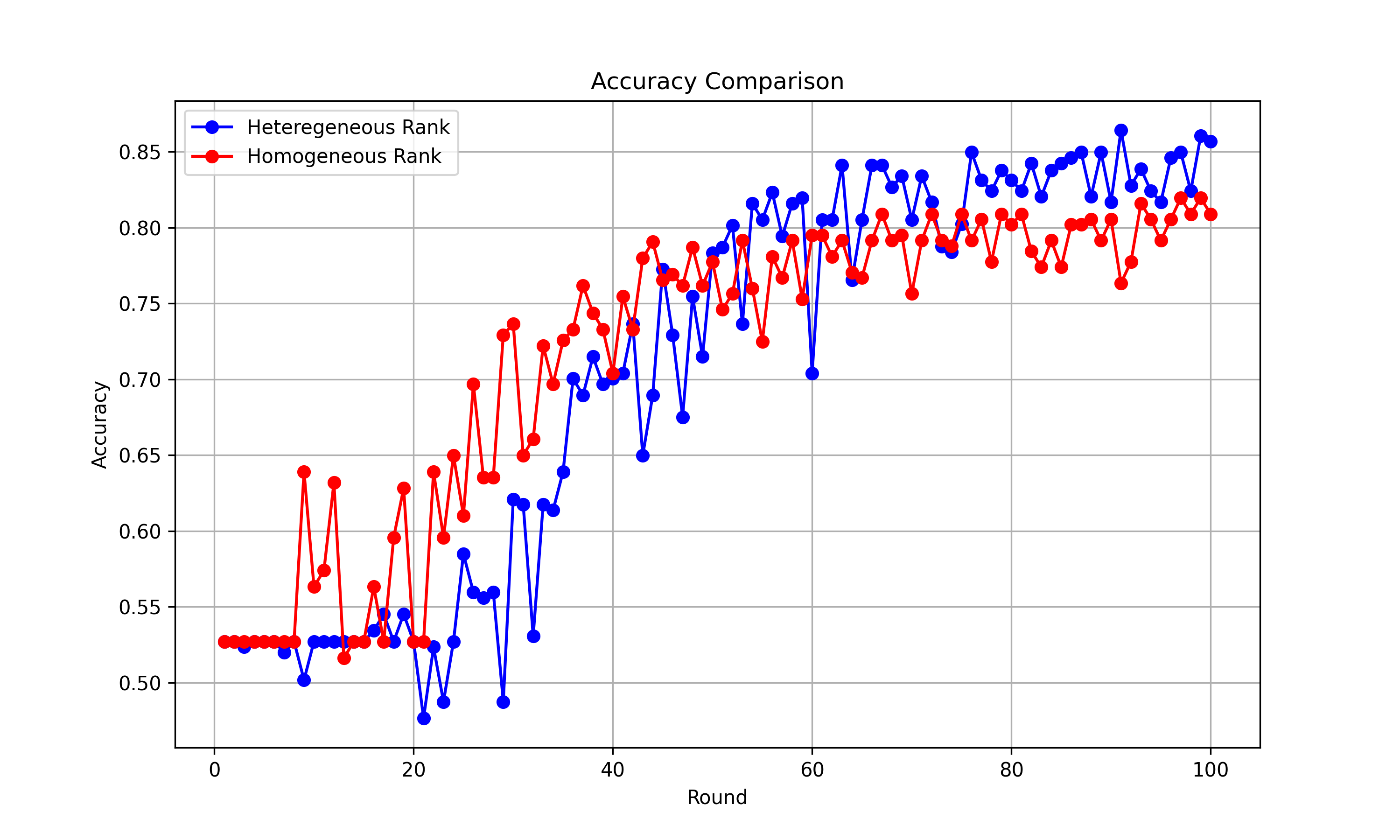}\label{fig:rte_2}} \\

         \vspace{-13pt}
        
    \subfloat[Naive implementation and Homogeneous \textit{HLoRA}(QQP)]{\includegraphics[width=0.5\textwidth, height=0.32\linewidth]{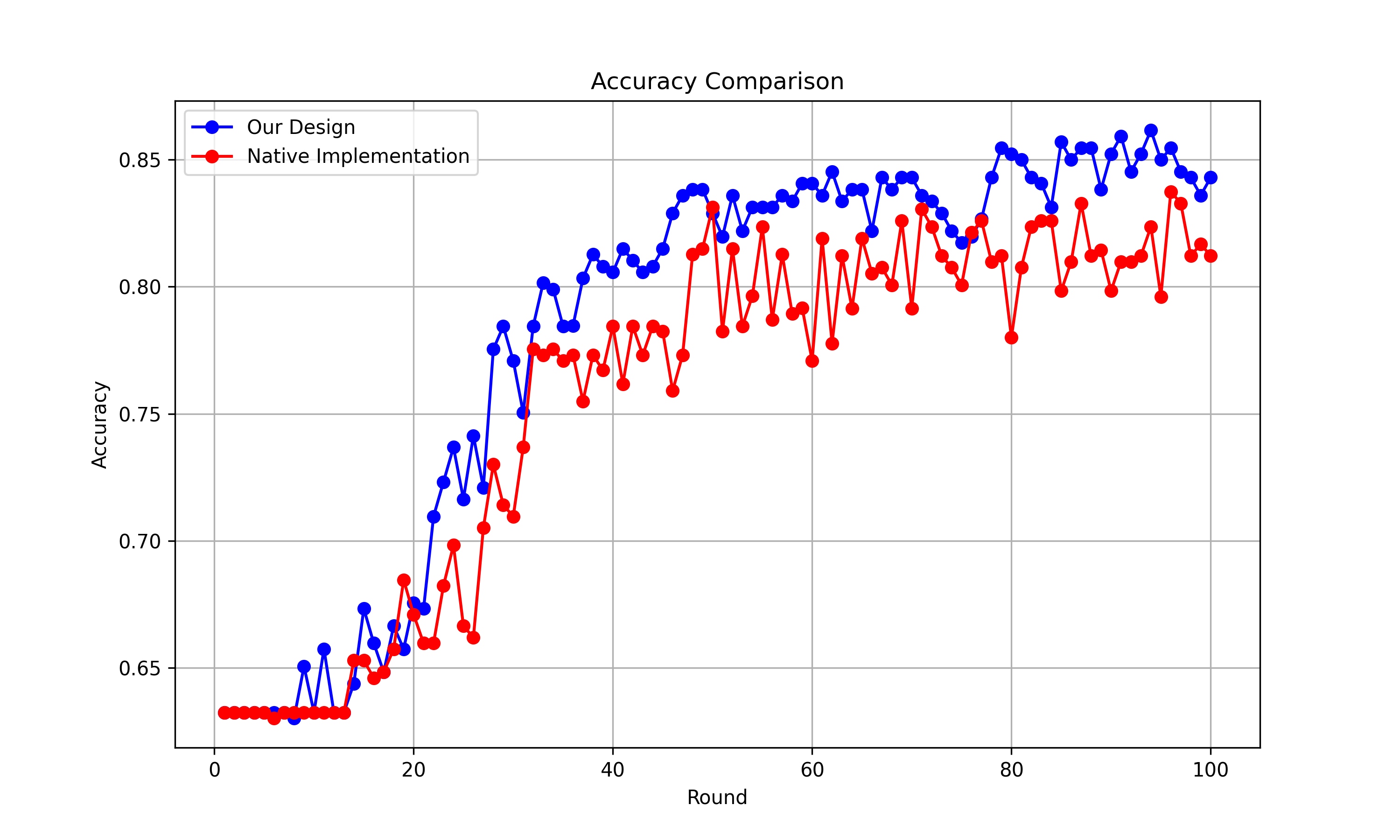}\label{fig:qqp_1}} 
        \subfloat[Homogeneous rank \textit{HLoRA} and Heterogeneous \textit{HLoRA}(QQP)]{\includegraphics[width=0.49\textwidth, height=0.32\linewidth]{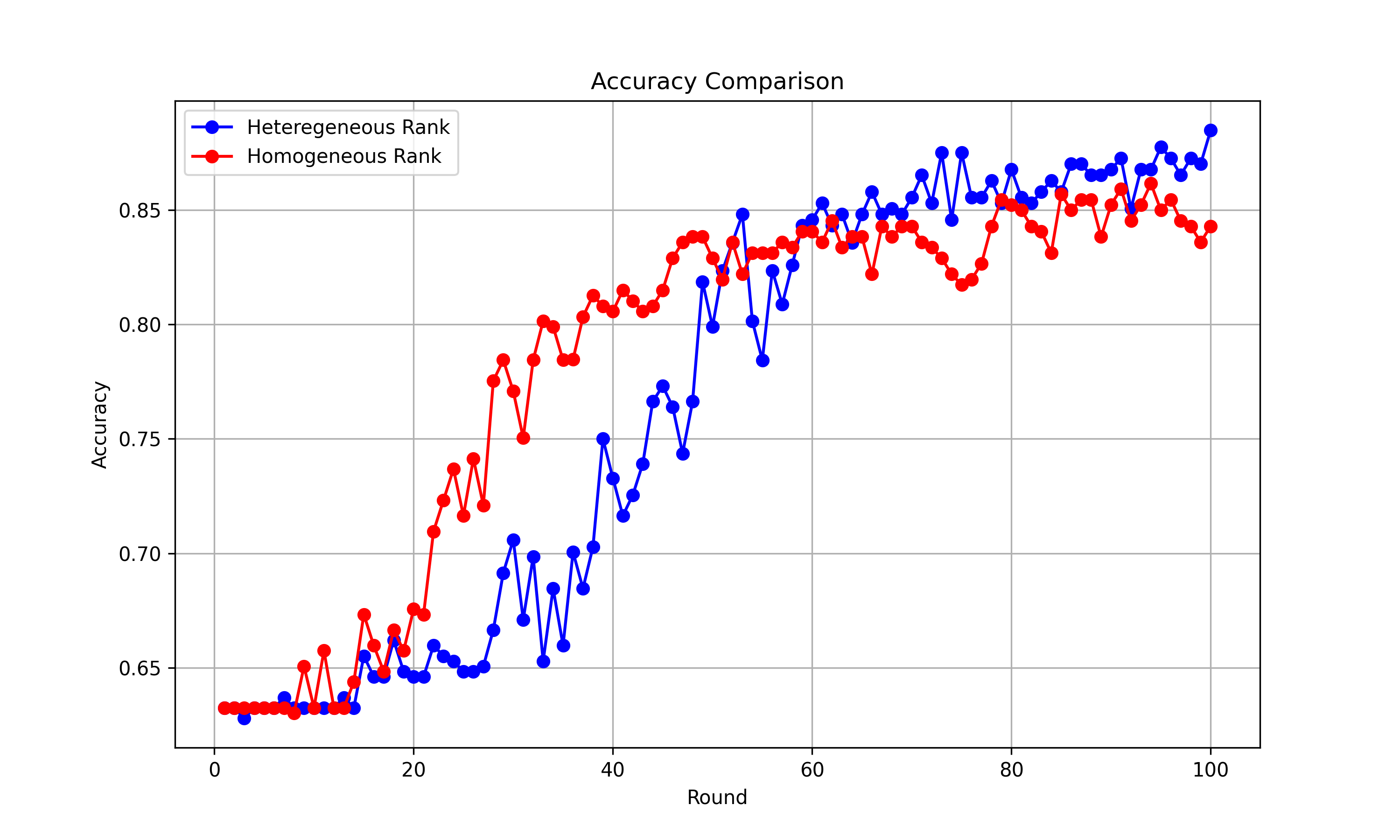}\label{fig:qqp_2}} 
    \caption{Comparative Performance Analysis of Federated LoRA Implementations. Sub-figure (a) shows the convergence speed and final performance of the naive implementation versus the reconstructed matrix re-decomposition with rank homogeneity, demonstrating faster convergence and higher ultimate performance in the latter. Sub-figure (b) compares the performance of reconstructed matrix re-decomposition with rank homogeneity against rank isomorphism, highlighting that while rank isomorphism converges more slowly, it achieves superior long-term accuracy. These comparisons underscore the impact of rank configuration on the efficacy of federated learning adaptations.}
    \vspace{-13pt}
    \label{fig:compare}
\end{figure*}

\begin{table*}[h!]
\centering
\begin{tabular}{c|c|c|c}
\toprule[1pt]
\textbf{Training Strategies} & \textsc{MRPC} & \textsc{RTE} & \textsc{QQP}\\
\toprule[1pt]
Centralised LoRA Fine-Tuning\cite{hu2021lora} & 90.2 & 87.4 & 91.6\\
Heterogeneous Rank Reconstruction & 87.1 & 86.1 & 88.4\\
Reconstruction Re-Decomposition (Rank Homogeneity) & 86.0 & 81.9 & 86.1 \\
Direct Application of LoRA (Naive Implementation) & 84.0& 78.3 & 83.7\\
\toprule[1pt]
\end{tabular}
\caption{Accuracy comparison for different kinds of training strategies based on various benchmarks.}
\label{tab:accuracy_comparison}
\vspace{-12pt}
\end{table*}

\section{Evaluation and Results Analysis}

\subsection{Prototype Implementation}
We implement our \textit{HLoRA} on top of \textit{Plato} \cite{li2023plato} and \textit{Pytorch}. \textit{Plato} is a federated learning framework that supports temporal simulation for both synchronous and asynchronous federated learning on a single device, such as a single GPU. Our evaluations for \textit{HLoRA} were performed on 6 NVIDIA GeForce RTX 4090 graphic cards.

\subsection{Evaluation Methodology}

\subsubsection{Model.} 
We evaluate our proposed \textit{HLoRA} based on the most popular model: RoBERTa-large \cite{liu2019roberta}.

\subsubsection{Datasets.}
We evaluate our proposed \textit{HLoRA} based on three datasets: Microsoft Research Paraphrase Corpus (MRPC)\cite{mrpc}, the Quora Question Pairs (QQP)\cite{qqp}, and the Recognizing Textual Entailment (RTE)\cite{rte}. 
Since we rely on non-IID distribution \cite{gliwaSAMSumCorpusHumanannotated2019, hsuMeasuringEffectsNonIdentical2019,liuRecentAdvancesFederated2023,liuUnderstandingLLMsComprehensive2024, liuWhatMakesGood2024} of data, we focus on classification tasks.
\begin{itemize}
    \item MRPC: The Microsoft Research Paraphrase Corpus task involves determining whether pairs of sentences in the corpus are semantically equivalent. 

    \item QQP: The Quora Question Pairs task focuses on identifying whether two questions asked on the Quora platform are duplicates, i.e., whether they have the same intent despite being phrased differently. 

    \item RTE: The Recognizing Textual Entailment task is designed to determine if a given hypothesis can be logically inferred from a provided premise.
\end{itemize}
These tasks, though individually distinct, collectively provide a comprehensive evaluation framework for assessing the model's performance across different dimensions of natural language understanding. By selecting these tasks, we ensure a rigorous evaluation of our \textit{HLoRA} approach and also provide an in-depth analysis of the evaluation results.

\textbf{Federated Setting.} We deploy 100 clients and sample 20 clients for each communication round. We deploy \textit{HLoRA} on a small-scale GPU cluster, including 6 NVIDIA GeForce RTX 4090 GPUs. We use these GPUs to simulate the server as well as all the clients.

\textbf{Hyper-Parameters.} We set the learning rate as $3e-4$ with local epoch E=2. For the hyperparameter in LoRA, we set $r=8$ for homogeneous rank setting and $r\in [2,8]$ for heterogeneous rank.

\textbf{Baseline.} We compare our proposed \textit{HLoRA} with other related approaches: (i) FedAvg\cite{mcmahan2017communication} and LoRA \cite{hu2021lora}, which simply and directly applied the LoRA into federated learning systems; (ii) \textit{HLoRA} with homogeneous rank; (iii) \textit{HLoRA} with heterogeneous rank.

\subsection{Results and Analysis}
Our experimental results provide a comprehensive comparison of different implementations of the federated \textit{LoRA} adaptation strategy. The results are illustrated through Fig. \ref{fig:compare} and Tab. \ref{tab:accuracy_comparison} that captures the performance variations under different conditions.

Fig. \ref{fig:compare} illustrates the performance comparison between the naive implementation of federated \textit{LoRA} (where no rank adaptation is applied) and our proposed method using reconstructed matrix re-decomposition with homogeneous rank and heterogeneous rank. It is evident from the figure that the reconstructed matrix in the re-decomposition method not only converges faster but also achieves superior performance by the end of the training process, which means that our training strategy can achieve the target accuracy using less training rounds compared to the baselines.

\subsubsection{Naive implementation vs Homogeneous rank \textit{HLoRA}:}
The naive implementation of federated \textit{LoRA}, as depicted in the figures, tends to lower accuracy and converge at a slower rate, which can be attributed to the introduction of bias as shown in Formula \ref{eq:bias_ab}. This bias results from the incompatibility of the LoRA adaptor's features with the Fedavg algorithm, leading to inefficiencies and slower learning. In contrast, our proposed method mitigates this issue by reconstructing the \textit{LoRA} adaptors through re-decomposition, ensuring effective aggregation. This approach effectively reduces the bias, thereby facilitating faster convergence.

\subsubsection{Heterogeneous rank \textit{HLoRA} vs Homogeneous rank \textit{HLoRA}:}
Fig. (\ref{fig:mrpc_2}), (\ref{fig:rte_2}) and (\ref{fig:qqp_2}) contrasts the performance of \textit{HLoRA} with rank homogeneity against an implementation with heterogeneous rank. Although heterogeneous rank converges more slowly, it ultimately outperforms the homogeneous rank approach in terms of final model accuracy.
In the experiments setting, the heterogeneous rank was set to take values ranging from 2 to 8, while the rank of isomorphic rank was fixed at 8. This means that heterogeneous ranks could only have a smaller rank, but achieve a higher accuracy rate.
This is due to the fact that not all stages in the fine-tuning process produce high-dimensional updates, and if the rank is large and the dimension of the update is small, there may be redundant parameters to update, which can lead to overfitting. However, the use of heterogeneous rank can avoid overfitting to some extent, thus improving the accuracy of the final model.

The comparative accuracies under different training settings are summarized in Tab. \ref{tab:accuracy_comparison}. This table reveals that there are still losses in the distributed setting compared to centralized fine-tuning, so centralized fine-tuning should still be taken wherever possible. However, among the federated strategies, the heterogeneous rank reconstruction approach performs best, followed by the homogeneous rank reconstruction, with the naive implementation lagging behind. These results underscore the importance of tailored rank strategies in enhancing the effectiveness of federated learning models under distributed conditions.

\subsubsection{Discussion}

The observed results demonstrate the critical role of rank adaptation in federated learning environments. By modifying the rank of adaptation matrices according to the heterogeneity of client capabilities, our proposed methods significantly outperform the naive implementation, which does not consider rank discrepancies among clients. The slower convergence rate of the rank isomorphism approach compared to rank homogeneity suggests a trade-off between initial learning speed and long-term model performance, which merits further investigation.


\section{Conclusion and Future Works} 
In this study, we explored the federated fine-tuning of Large Language Models (LLMs) tailored to the inherent system and data heterogeneity through our proposed framework. We demonstrate that our approach is not only feasible but also surpasses the conventional implementation of Layerwise Relevance Propagation (LoRA) in terms of computational efficiency and overall performance. Our findings prompt several intriguing research questions. Notably, within specific settings that permit the assignment of distinct ranks to clients, what would be the optimal method for distributing these ranks to enhance convergence and performance outcomes? Currently, our system assigns these ranks randomly among clients; however, whether a targeted assignment strategy could improve the heterogeneous performance of LoRA warrants further exploration.

\bibliography{aaai25}

\end{document}